\documentclass[sn-mathphys,Numbered]{sn-jnl}

\usepackage{graphicx}%
\usepackage{multirow}%
\usepackage{amsmath,amssymb,amsfonts}%
\usepackage{amsthm}%
\usepackage{mathrsfs}%
\usepackage[title]{appendix}%
\usepackage{xcolor}%
\usepackage{textcomp}%
\usepackage{manyfoot}%
\usepackage{booktabs}%
\usepackage{algorithm}%
\usepackage{algorithmicx}%
\usepackage{algpseudocode}%
\usepackage{listings}%
\usepackage{biblatex}
\usepackage{caption}
\usepackage{multirow}

\theoremstyle{thmstyleone}%
\theoremstyle{thmstyletwo}%

\theoremstyle{thmstylethree}%

\begin{document} 

\title[Article Title]{{Diffuse X-ray Explorer: a high-resolution X-ray spectroscopic sky surveyor on the China Space Station}
}


\author *[1]{\fnm{Hai} \sur{Jin}}\email{jinhai@tsinghua.edu.cn}
\author*[1]{\fnm{Junjie} \sur{Mao}}\email{jmao@tsinghua.edu.cn}

\author[2]{\fnm{Liubiao} \sur{Chen}}\email{chenliubiao@mail.ipc.ac.cn}
\author[1]{\fnm{Naihui} \sur{Chen}}\email{cnh21@mails.tsinghua.edu.cn}
\author[1]{\fnm{Wei} \sur{Cui}}\email{cui@tsinghua.edu.cn}
\author[3]{\fnm{Bo} \sur{Gao}}\email{bo\_f\_gao@mail.sim.ac.cn}
\author[4]{\fnm{Jinjin} \sur{Li}}\email{jinjinli@nim.ac.cn}
\author[5]{\fnm{Xinfeng} \sur{Li}}\email{ lixinfeng@csu.ac.cn}
\author[1]{\fnm{Jiejia} \sur{Liu}}\email{liujj21@mails.tsinghua.edu.cn}
\author[2]{\fnm{Jia} \sur{Quan}}\email{quanjia10@mail.ipc.ac.cn}
\author[1]{\fnm{Chunyang} \sur{Jiang}}\email{jiangcy22@mails.tsinghua.edu.cn}
\author[6]{\fnm{Guole} \sur{Wang}}\email{glwang@bao.ac.cn}
\author[6]{\fnm{Le} \sur{Wang}}\email{wangle@nao.cas.cn}
\author[1]{\fnm{Qian} \sur{Wang}}\email{qianwang0304@mail.tsinghua.edu.cn}
\author[1]{\fnm{Sifan} \sur{Wang}}\email{wsf18@mails.tsinghua.edu.cn}
\author[5]{\fnm{Aimin} \sur{Xiao}}\email{Am.Xiao@csu.ac.cn}
\author[7]{\fnm{Shuo} \sur{Zhang}}\email{shuozhang@shanghaitech.edu.cn}

\affil*[1]{\orgdiv{Department of Astronomy}, \orgname{Tsinghua University},  \city{Beijing}, \postcode{100084}, \country{China}}

\affil[2]{\orgdiv{Technical Institute of Physics and Chemistry}, \orgname{Chinese Academy of Sciences},  \city{Beijing}, \postcode{100190},  \country{China}}

\affil[3]{\orgdiv{Shanghai Institute of Microsystem and Information Technology},\orgname{Chinese Academy of Sciences},  \city{Shanghai}, \postcode{200050},  \country{China}}

\affil[4]{\orgdiv{National Institute of Metrology}, \city{Beijing}, \postcode{102200},  \country{China}}

\affil[5]{\orgdiv{Technology and Engineering Center for Space Utilization}, \orgname{Chinese Academy of Sciences}, \city{Beijing}, \postcode{100094},  \country{China}}

\affil[6]{\orgdiv{National Astronomical Observatories}, \orgname{Chinese Academy of Sciences}, \city{Beijing}, \postcode{100101},  \country{China}}

\affil[7]{\orgdiv{ShanghaiTech University}, \city{Shanghai}, \postcode{201210},  \country{China}}


\abstract{DIffuse X-ray Explorer (DIXE) is a proposed high-resolution X-ray spectroscopic sky surveyor on the China Space Station (CSS). DIXE will focus on studying hot baryons in the Milky Way. Galactic hot baryons like the X-ray emitting Milky Way halo and eROSITA bubbles are best observed in the sky survey mode with a large field of view. DIXE will take advantage of the orbital motion of the CSS to scan a large fraction of the sky. High-resolution X-ray spectroscopy, enabled by superconducting microcalorimeters based on the transition-edge sensor (TES) technology, will probe the physical properties (e.g., temperature, density, elemental abundances, kinematics) of the Galactic hot baryons. This will complement the high-resolution imaging data obtained with the eROSITA mission. Here we present the preliminary design of DIXE. The payload consists mainly of a detector assembly and a cryogenic cooling system. The key components of the detector assembly are a microcalorimeter array and frequency-domain multiplexing readout electronics. To provide a working temperature for the detector assembly, the cooling system consists of an adiabatic demagnetization refrigerator and a mechanical cryocooler system.}




\maketitle

\section{Introduction}\label{sec1}
The galactic ecosystem is one of the major subjects in astronomy that requires further studies. Hot baryons, best observed in the X-ray band, play an important role in the galactic ecosystem. These hot baryons trace the feedback processes from supermassive black holes (SMBH) and stars, which have a profound impact on the formation and evolution of galaxies. Nonetheless, the detailed physics of both feedback processes is poorly understood. 

The Milky Way, our mother galaxy, is a critical template for understanding the galactic ecosystem in general. The eROSITA instrument aboard the Russian-German Spectrum-Roentgen-Gamma (SRG)\cite{Predehl21} mission has greatly updated our understanding of hot baryons in the Milky Way \cite{Predehl20}. Compared to the previous X-ray all-sky map obtained with the ROSAT mission in the 1990s \cite{Snowden95}, the one obtained with eROSITA has unprecedented spatial resolution and sensitivity. For instance, the pair of eROSITA bubbles above and below the galactic plane, covering over $10^4$ square degrees in the sky, seems to be originated from the Galactic center \cite{Predehl20}. The exact origin of eROSITA bubbles is still debated though \cite{Predehl20, Yang22, Gupta23}. 

The eROSITA instrument adopt Charge-Coupled Device (CCD) as its detector, which provides an energy resolution of $\sim$60 eV at 0.5 keV and $\sim$140 eV at 6.4 keV \cite{Predehl21}, which is insufficient to resolve characteristic emission lines from cosmically abundant elements (e.g., C, N, O, Ne, Mg, Fe) in the Galactic hot baryons. For instance, the resonance, intercombination, and forbidden lines from the He-like triplet of O {\sc vii} are at the rest-frame energy of 574 eV, 567 eV, and 561 eV. On the one hand, the line ratios of He-like triplets are sensitive to the temperature and density of the plasma \cite{Porquet10}. On the other hand, the line ratios can also be used to distinguish collisional
ionized equilibrium plasmas and photoionized plasmas \cite{Mao19}. Moreover, combined with the Lyman series, the role of the solar wind charge exchange (SWCX) process can also be quantified \cite{Mao22}.

\section{Scientific Objectives}\label{sec2}
The main goal of DIXE is to probe the physical properties (e.g., temperature, density, elemental abundances, kinematics) of the Galactic hot baryons. Galactic hot baryons are in fact the foreground of any extragalactic X-ray observations. This adds to the scientific return of DIXE. 

Within the solar system, the SWCX process can produce time-variable X-ray foreground emission \cite{Kuntz19}. The hot solar wind can pick up electrons from the neutral medium in the solar system, capturing them into excited energy levels rarely populated in other mechanisms (e.g., collisional ionization equilibrium, photoionization equilibrium). The unstable excited energy level is de-populated by line emissions. The characteristic features of SWCX are the enhanced forbidden line of the He-like triplet with respect to the resonance line and the enhanced high-order Lyman series line with respect to the Ly$\alpha$ line \cite{Mao22, Gu23}. High-resolution X-ray spectroscopy is the key to quantifying the role of SWCX in such kind of X-ray foreground emission.  

The solar system resides in the Local Hot Bubble (LHB) with a size scale of ~100 pc \cite{Zucker22}, which also produces bright X-ray emission. Although LHB is irregular in shape \cite{Zucker22}, it is thought to have a uniform temperature of $\sim$0.1 keV \cite{Sanders2001}. Its intensity, unattenuated by the Galactic absorption, varies across the sky. Its elemental abundances are largely unknown. Some fractions of O, Mg, Si, and Fe might be locked up in dust \cite{Rogantini22, Psaradaki23}, but the exact dust depletion fraction is poorly constrained. 

The Milky Way Halo (MWH) is postulated as a spherical symmetric halo to the first-order approximation \cite{Henley12, Gupta12}. At least some (if not all) fractions of MWH have a temperature of $\sim0.1-0.2$ keV \cite{McCammon02}. Similar to the LHB, the metallicity of the MWH is largely unknown due to the lack of high-resolution X-ray spectra. Moreover, a hotter component with a temperature of $\sim$0.7 keV accompanied by a super-solar Ne/O abundance ratio has also been suggested \cite{Das19}. Such a component is not expected within the framework of a single-temperature hot Galactic halo. If this component is verified with DIXE observations, we will further investigate if it originates from M dwarf stars in the Galactic disk \cite{Wulf19} or if it is part of a two-temperature Galactic halo \cite{Das19}. 

Between LHB and MHW, large-scale (over a few hundreds of square degrees) hot baryons are also observable to DIXE. As mentioned in the introduction (Section~\ref{sec1}), the origin of the eROSITA bubble is still debated \cite{Predehl20, Yang22, Gupta23}. Some argue it originates from the past activity of Milky Way’s SMBH (Sgr A$\star$) \cite{Yang22}, while others favor a stellar feedback origin \cite{Gupta23}. The confusion mainly arises from the relatively poor measurement of elemental abundances as well as the emission measure distribution as a function of temperature. In addition, DIXE can constrain the kinematics if the bubbles have a significant Doppler shift due to their motion. According to the simulation by \cite{Yang22}, the bubble has a relatively hot component traced by the Fe {\sc xxv} He-like triplets (rest-frame energy: $\sim$6.7 keV) and is expanding with a velocity of 2000 ${\rm km~s^{-1}}$ (i.e., $\Delta E=45 eV $ at 6.7 keV), DIXE observation can easily verify this theoretical prediction. Another example is the Cygnus loop, which is a nearby supernova remnant (SNR) where charge exchange features \cite{Katsuda11} and the ``low-abundance problem” \cite{Uchida19} were suggested for some local structures. A global high-resolution X-ray spectroscopic view of the entire SNR is still lacking. 

\section{Key Specifications}\label{sec3}

\begin{figure}
    \centering
    \includegraphics[width=1\linewidth]{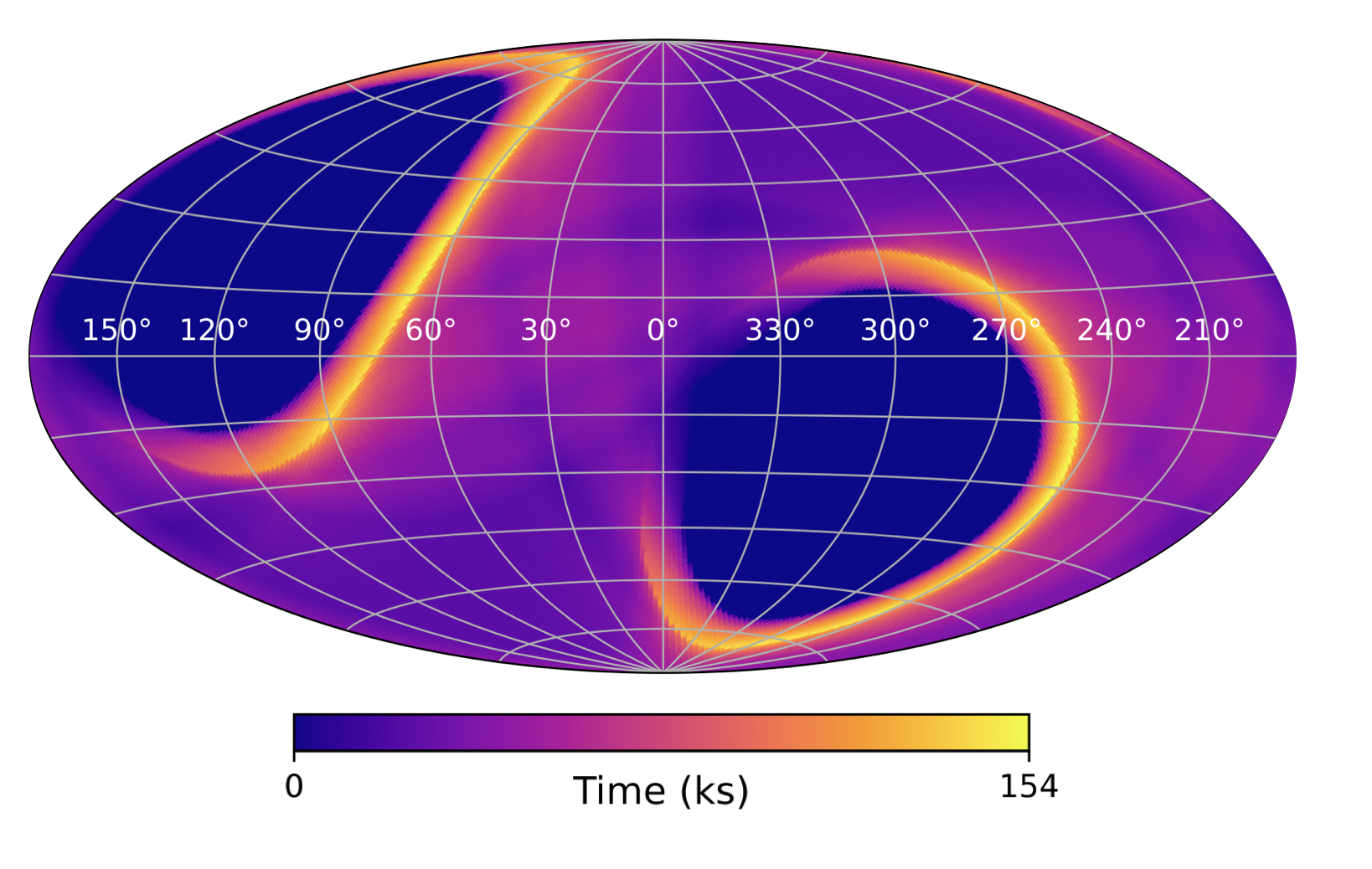}
    \caption{Expected 1 yr DIXE sky survey exposure map shown in Galactic coordinates, using a Hammer–Aitoff projection.}
    \label{fig1}
\end{figure}

Driven by different scientific goals, the key design specifications of DIXE differ from those of eROSITA (Tab~\ref{tab:my_table1}). Equipped with microcalorimeters, DIXE will achieve an energy resolution of 6 eV at 0.6 keV, which makes it possible to resolve key diagnostic lines like He-like triplets of O {\sc vii} to advance our knowledge of the physical properties of the hot baryons in the Milky Way. With a collimated $10^{\circ}\times10^{\circ}$ field of view and following the orbit of CSS, DIXE will cover a large fraction of the sky (Fig~\ref{fig1}) in an efficient way. The microcalorimeters are based on transition-edge sensor (TES) technology and need to be cooled to below 100 mK to realize their superior energy resolution. The cooling is provided by a system consisting of the adiabatic demagnetization refrigerator (ADR) and mechanical cryocoolers. The microcalorimeter array is read out with frequency-domain multiplexing (FDM) electronics. The key technologies developed for DIXE will pave the way for the development of the Hot Universe Baryon Surveyor (HUBS) mission~\cite{Cui20}.

\begin{table}
\caption{Comparing the design of eROSITA \cite{Predehl21}, DIXE and HUBS \cite{Cui20}. The scientific goals are Large-scale structure and cosmology for eROSITA \cite{Predehl21}, Milky Way hot baryons for DIXE, and Cosmic hot baryons for HUBS \cite{Cui20}. The design of these missions differs mainly driven by their distinguished scientific goals. \label{tab:my_table1}}
\centering
\begin{tabular}{c|cccc}
\noalign{\smallskip}\hline\noalign{\smallskip}
Mission & eROSITA & DIXE & HUBS \\
\noalign{\smallskip}\hline\noalign{\smallskip}
Platform & Satellite & CSS & Satellite \\
\noalign{\smallskip}\hline\noalign{\smallskip}
Observing mode & Scan + Pointing & Scan & Pointing \\
\noalign{\smallskip}\hline\noalign{\smallskip}
FOV & $1^{\circ}\times1^{\circ}$ & $10^{\circ}\times10^{\circ}$ & $1^{\circ}\times1^{\circ}$ \\ 
    & (focusing optics) & (collimator) & (focusing optics)  \\
\noalign{\smallskip}\hline\noalign{\smallskip}
Energy band & $0.3-10$ keV & $0.1-10$ keV & $0.1-2$ keV \\
\noalign{\smallskip}\hline\noalign{\smallskip}
Detector & CCD & TES microcalorimeter & TES microcalorimeter \\
     & & ($10\times10$ array) & ($60\times60$ array) \\
\noalign{\smallskip}\hline\noalign{\smallskip}
Energy resolution &  60 eV@0.5 keV & 6 eV @0.6 keV & 2 eV 0.6 keV \\
\noalign{\smallskip}\hline\noalign{\smallskip}
Launch time & 2019 & 2027 (expected) & 2031 (expected) \\
\noalign{\smallskip}
\hline
\end{tabular}
\end{table}



\section{Preliminary Designs}
\subsection{Overview}
\begin{figure}
    \centering
    \includegraphics[width=1\linewidth]{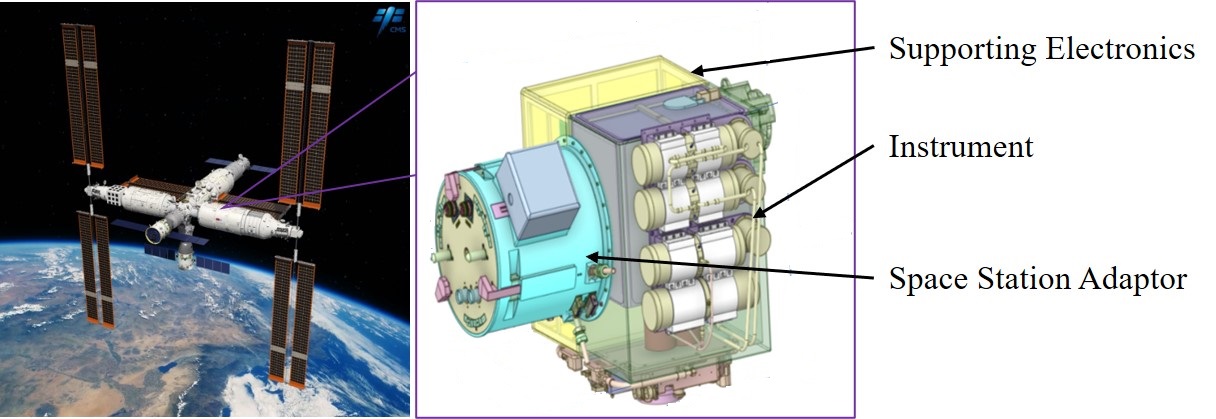}
    \caption{The DIXE experiment: ({\it left}) Artist's view of the China Space Station. Credit: China Manned Space Agency (https://www.cmse.gov.cn/); ({\it right}) schematics of the payload }
    \label{fig2}
\end{figure}

The DIXE payload will be installed on the external hanging point of the CSS, as shown in Fig.~\ref{fig2} (left), with the direction of pointing fixed. It scans the sky as the CSS orbits the Earth. The DIXE payload consists of three parts, as also shown in Fig.~\ref{fig2} (right), including the instrument, supporting electronics, and space station adapter. Through the space station adapter electric power and cooling are provided by the CSS. The electronics include modules for multiplexing readout, detector temperature control, magnet current control, cryocooler control, and other necessary functions of the payload. 

\subsection{Payload Design}
The field of view is mechanically collimated to $10^{\circ}\times10^{\circ}$ (at FWHM). An array of TES-based microcalorimeters is chosen for the detector, providing an energy resolution of 6 eV at 0.6 keV. The array is $10\times10$ in size, covering roughly an area of 1 cm$^2$. The pixels are divided into 4 quadrants, with each read out with an FDM module (i.e., the multiplexing factor is 25). The detector is mounted on a cold plate whose temperature is maintained below 100 mK. Cooling is provided by the combination of mechanical cryocoolers and an adiabatic demagnetization refrigerator (ADR). 

\begin{figure}
    \centering
    \includegraphics[width=1\linewidth]{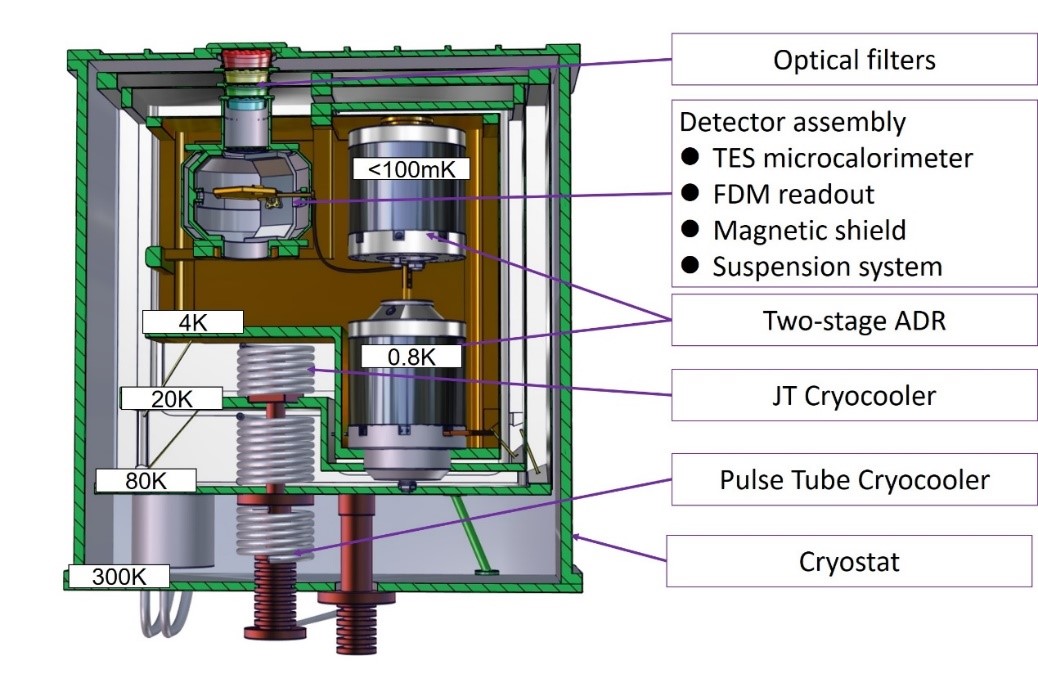}
    \caption{Preliminary design of DXIE instrument}
    \label{fig:pre.design}
\end{figure}

Fig.~\ref{fig:pre.design} shows the inside layout of the payload. Functionally, the payload can be divided into three parts: (1) detector assembly;(2) cooling system; and (3) adaptor interface to CSS. In the following, we describe the first two in more detail.

\subsubsection{Detector Assembly}
The detector assembly is designed as a fully integrated unit \cite{LJJ_DetAss}. The outermost layer of the unit is magnetic shielding, which serves to shield the residual magnetic field of the ADR and geomagnetic field, to minimize adverse effects on the microcalorimeters. The goal is to reduce the magnetic field to a level below $1\mu T$ near the detector array. This can be accomplished by adopting a two-layer design: an outer Cryoperm layer and an inner Nb layer, based on the results of simulations \cite{LJJ_DetAss}. Above the detector, there are five optical blocking filters installed at different temperature stages, to minimize the degradation in the energy resolution of the detector due to shot noises mainly caused by incident radiation at optical/IR wavelengths. 

At the center of the detector assembly lies the TES microcalorimeter array that is surrounded by the cold stage of the FDM system. The detector mounting plate is connected to the ADR thermally and is suspended by a Kevlar support system off the 4K stage. For DIXE microcalorimeters, the TES is based on Mo/Cu bilayer film material, and the absorber is made of Au and Bi. The absorber of each pixel needs to be 1 mm in size, in order to reach a total detector area of 1 cm$^2$. As designed, a multiplexing factor of 25:1 is required for the FDM system under development. The input SQUIDs and LC resonators are placed near the detector array on the same cold plate, with the signals initially amplified by a SQUID array at the 4K stage and then passed along to the warm electronics.

\subsubsection{Cooling System}

Due to the modular design, the upgoing transport vehicle (to CSS) imposes severe size restrictions on the DIXE payload. At present, the maximum dimensions are estimated to be 530~mm$\times$600~mm$\times$610~mm, so the cooling system must be very compact. As a preliminary design, two pulse tubes are used to cool the system to 80~K and then to 20~K, from which a JT cooler is used to cool the system further down to 4~K. Radiation shields are installed at these temperature stages to cut down heat load on the cold plate due to thermal radiation from the outer jacket of the vacuum dewar(Fig.3). 

At 4 K, the detector assembly and the ADR, which reside inside the 4 K shield, are isolated from the rest of the system through the opening of a heat switch and are further cooled down  below 100 mK by the ADR. A two-stage design is baselined for the ADR, with the GGG (Gadolinium Gallium Garnate) stage providing cooling to about 0.8 K and the FAA (Ferric Ammonium Alum) stage to its operating temperature (below 100 mK). 

\begin{table}
\caption{Design of the DIXE readout, ADR mK cooler, and mechanical cooling sytems. \label{tab:design para.}}
\centering
\begin{tabular}{c|c|ccc}
\noalign{\smallskip}\hline\noalign{\smallskip}
System & Parameter & Value \\
\noalign{\smallskip}\hline\noalign{\smallskip}
\multirow{2}{*}{Readout} & Multiplexing type & FDM \\
    & Multiplexing factor & 25 \\
\noalign{\smallskip}\hline\noalign{\smallskip}
\multirow{3}{*}{ADR mK cooler} & Minimum temperature & $50$ \\
    & Hold time & $> 9$ hours \\
    & Duty cycle & $>90$~\% \\
\noalign{\smallskip}\hline\noalign{\smallskip}
\multirow{3}{*}{Mechanical cooler} & 4 K stage cooling power & 50 mW @ 4K \\
    & 80 K stage cooling power & 10 mW @ 80 K \\
    & power consumption & $<640$ W (AC) \\
\noalign{\smallskip}
\hline
\end{tabular}
\end{table}


\section{Technical Development}

\subsection{Detector}
\begin{figure}
    \centering
    \includegraphics[width=0.5\linewidth]{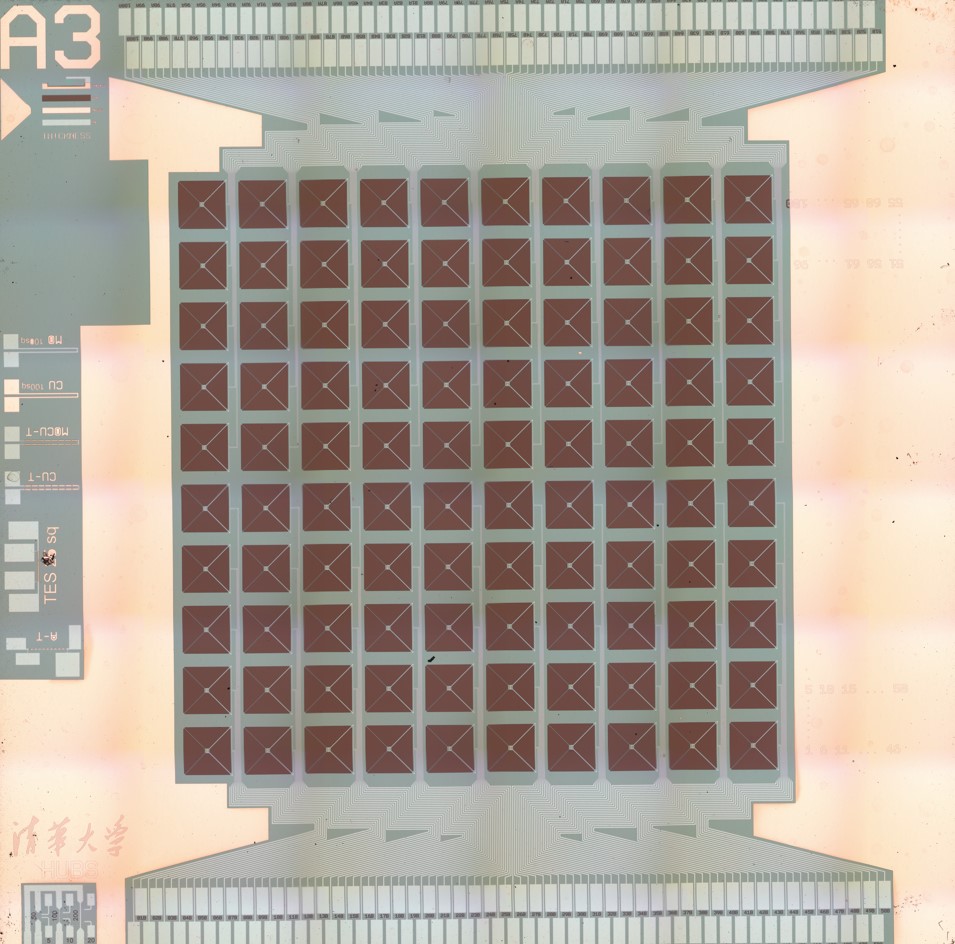}
    \caption{Photo of a 10x10 TES test array. This shows one of the designs under consideration for slow devices. }
    \label{fig:TES MEMS}
\end{figure}

Detector development is ongoing~\cite{WANG2022100027}, with a number of test arrays fabricated with and without absorbers. Fig.~\ref{fig:TES MEMS} shows a $10\times10$ TES array. Particular attention has been paid to the control of thermal conductance between a pixel and the substrate through the perforation of the SiN$_x$ device-supporting film and also to the fabrication of large absorbers. For DIXE, slow devices are preferred, which would reduce the bandwidth requirement on the multiplexing readout electronics.

\subsection{Multiplexing Readout Electronics}
 
\begin{figure}
    \centering
    \includegraphics[width=0.75\linewidth]{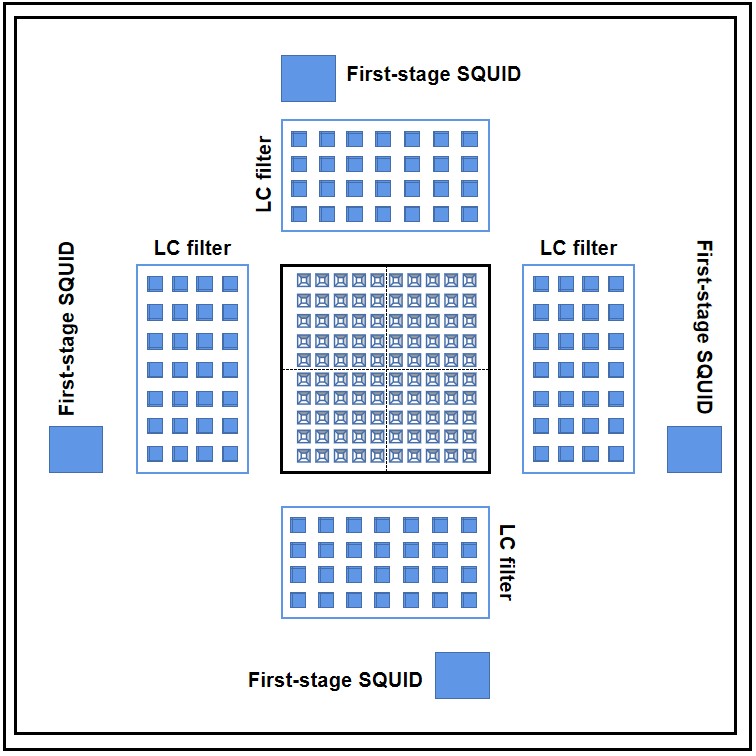}
    \caption{The FDM readout cold part prototype, with LC filter and SQUID surrounding TES array.}
    \label{fig:FDM}
\end{figure}

 Both FDM and TDM schemes are under development for DIXE, although the baseline design is based on FDM. Fig.~\ref{fig:FDM} shows a layout of the detector plane, where the microcalorimeter array and the front-end FDM readout modules are located. For DIXE, four FDM modules are needed, with each reading out 25 pixels. Progress are being made in the fabrication of LC filters and 2-stage SQUIDs. The first version of the LC filter array is successfully fabricated and its characterization is underway. The linewidth of the inductors is 2 $\mu m$ and the dielectric constant of the capacitor is 11. Both input SQUIDs and SQUID arrays are fabricated and tested. The readout noise of the input SQUID is less than 6 $pA/\sqrt{Hz}$, while the gain of the SQUID array is larger than 2000. The detailed results will be presented in a future publication.
As for TDM, critical components are being developed~\cite{SIMIT_TDM}. Recent progress is reported in~\cite{SIMIT_LTD}.
  
\subsection{Mechanical Cooler}
\begin{figure}
    \centering
    \includegraphics[width=1\linewidth]{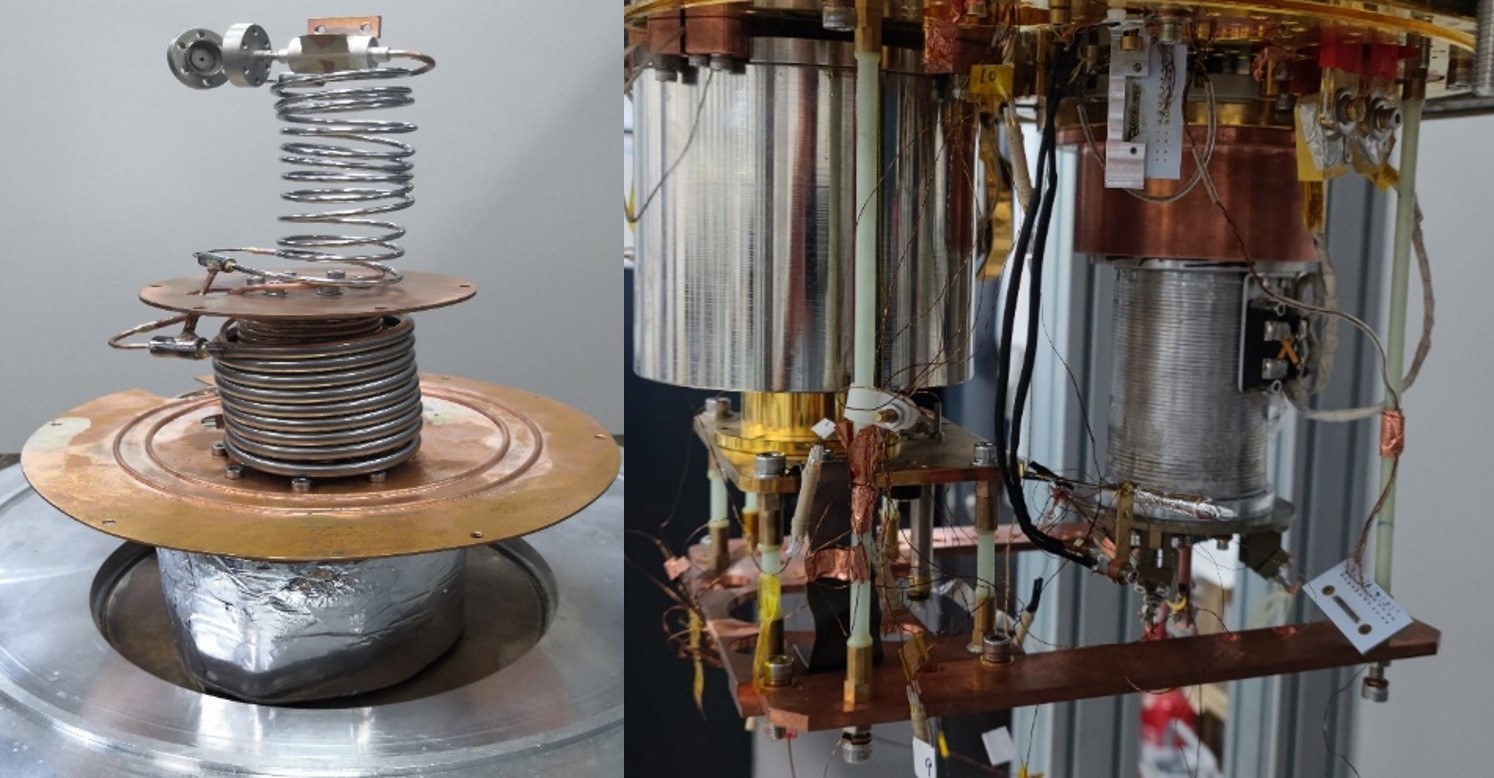}
    \caption{Photos of prototypes: ({\it left}) JT cooler; ({\it right}) two-stage ADR.}
    \label{fig:JT and ADR}
\end{figure}

A high-frequency pulse-tube prototype has been constructed. The tests show that it has reached a minimum temperature of 2.5 K, and provided a cooling power of 26 mW at 4.2 K \cite{2023YangB}. 
The JT cooler is under development. Fig.~\ref{fig:JT and ADR} (left) shows the prototype. The test results show that it is capable of providing 100.2~mW cooling power at 4.1~K, with 500~W input power.

\subsection{Adiabatic Demagnetization Refrigerator}
A two-stage ADR prototype has been constructed with Chromium Potassium Alum (CPA) used for the colder stage (instead of the FAA). The tests show that it is capable of reaching a temperature of about 44 mK, and can be held at 100 mK for about 3 hours \cite{JIANG2023}. For DIXE, the hold time is expected to be 9 hours or longer. In the meantime, a number of FAA salt pills have been made for the baseline FAA+GGG system and tested. Integration with the cryocooler system is planned for the testing of the integrated cooling system.

\section{Project Status}

DIXE was proposed to the China Manned Space Agency in 2019, in response to a solicitation for science experiments to be conducted on the CSS, and is among the four astronomy programs remaining for final approval. A start of the operation in 2027 is envisioned. A preliminary design of the payload has been completed. 
The development of key technologies, including TES microcalorimeter, FDM readout electronics, cryocooler and ADR, is progressing.

\backmatter

\bmhead{Acknowledgments}

We wish to thank Dr. Dan McCammon and all members of the DIXE collaboration team for useful discussion. This work was supported in part by the Ministry of Science and Technology of China through Grant 2022YFC2205100, by China National Space Administration (CNSA) through a technology development grant, and by the National Natural Science Foundation of China through Grants 11927805, 11803014 and 12220101004.


\begin{refcontext}[sorting=none]
\printbibliography

@ARTICLE{Predehl21,
       author = {{Predehl}, P. and {Andritschke}, R. and {Arefiev}, V. and {Babyshkin}, V. and {Batanov}, O. and {Becker}, W. and {B{\"o}hringer}, H. and {Bogomolov}, A. and {Boller}, T. and {Borm}, K. and {Bornemann}, W. and {Br{\"a}uninger}, H. and {Br{\"u}ggen}, M. and {Brunner}, H. and {Brusa}, M. and {Bulbul}, E. and {Buntov}, M. and {Burwitz}, V. and {Burkert}, W. and {Clerc}, N. and {Churazov}, E. and {Coutinho}, D. and {Dauser}, T. and {Dennerl}, K. and {Doroshenko}, V. and {Eder}, J. and {Emberger}, V. and {Eraerds}, T. and {Finoguenov}, A. and {Freyberg}, M. and {Friedrich}, P. and {Friedrich}, S. and {F{\"u}rmetz}, M. and {Georgakakis}, A. and {Gilfanov}, M. and {Granato}, S. and {Grossberger}, C. and {Gueguen}, A. and {Gureev}, P. and {Haberl}, F. and {H{\"a}lker}, O. and {Hartner}, G. and {Hasinger}, G. and {Huber}, H. and {Ji}, L. and {Kienlin}, A. v. and {Kink}, W. and {Korotkov}, F. and {Kreykenbohm}, I. and {Lamer}, G. and {Lomakin}, I. and {Lapshov}, I. and {Liu}, T. and {Maitra}, C. and {Meidinger}, N. and {Menz}, B. and {Merloni}, A. and {Mernik}, T. and {Mican}, B. and {Mohr}, J. and {M{\"u}ller}, S. and {Nandra}, K. and {Nazarov}, V. and {Pacaud}, F. and {Pavlinsky}, M. and {Perinati}, E. and {Pfeffermann}, E. and {Pietschner}, D. and {Ramos-Ceja}, M.~E. and {Rau}, A. and {Reiffers}, J. and {Reiprich}, T.~H. and {Robrade}, J. and {Salvato}, M. and {Sanders}, J. and {Santangelo}, A. and {Sasaki}, M. and {Scheuerle}, H. and {Schmid}, C. and {Schmitt}, J. and {Schwope}, A. and {Shirshakov}, A. and {Steinmetz}, M. and {Stewart}, I. and {Str{\"u}der}, L. and {Sunyaev}, R. and {Tenzer}, C. and {Tiedemann}, L. and {Tr{\"u}mper}, J. and {Voron}, V. and {Weber}, P. and {Wilms}, J. and {Yaroshenko}, V.},
        title = "{The eROSITA X-ray telescope on SRG}",
      journal = {Astronomy and Astrophysics},
         year = 2021,
        month = mar,
       volume = {647},
          eid = {A1},
        pages = {A1},
          doi = {10.1051/0004-6361/202039313},
archivePrefix = {arXiv},
       eprint = {2010.03477},
 primaryClass = {astro-ph.HE},
       adsurl = {https://ui.adsabs.harvard.edu/abs/2021A&A...647A...1P}
}

@ARTICLE{Predehl20,
       author = {{Predehl}, P. and {Sunyaev}, R.~A. and {Becker}, W. and {Brunner}, H. and {Burenin}, R. and {Bykov}, A. and {Cherepashchuk}, A. and {Chugai}, N. and {Churazov}, E. and {Doroshenko}, V. and {Eismont}, N. and {Freyberg}, M. and {Gilfanov}, M. and {Haberl}, F. and {Khabibullin}, I. and {Krivonos}, R. and {Maitra}, C. and {Medvedev}, P. and {Merloni}, A. and {Nandra}, K. and {Nazarov}, V. and {Pavlinsky}, M. and {Ponti}, G. and {Sanders}, J.~S. and {Sasaki}, M. and {Sazonov}, S. and {Strong}, A.~W. and {Wilms}, J.},
        title = "{Detection of large-scale X-ray bubbles in the Milky Way halo}",
      journal = {Nature},
         year = 2020,
        month = dec,
       volume = {588},
       number = {7837},
        pages = {227-231},
          doi = {10.1038/s41586-020-2979-0},
archivePrefix = {arXiv},
       eprint = {2012.05840},
 primaryClass = {astro-ph.GA},
       adsurl = {https://ui.adsabs.harvard.edu/abs/2020Natur.588..227P}
}

@ARTICLE{Snowden95,
       author = {{Snowden}, S.~L. and {Freyberg}, M.~J. and {Plucinsky}, P.~P. and {Schmitt}, J.~H.~M.~M. and {Truemper}, J. and {Voges}, W. and {Edgar}, R.~J. and {McCammon}, D. and {Sanders}, W.~T.},
        title = "{First Maps of the Soft X-Ray Diffuse Background from the ROSAT XRT/PSPC All-Sky Survey}",
      journal = {Astrophysical Journal},
         year = 1995,
        month = dec,
       volume = {454},
        pages = {643},
          doi = {10.1086/176517},
       adsurl = {https://ui.adsabs.harvard.edu/abs/1995ApJ...454..643S}
}

@ARTICLE{Yang22,
       author = {{Yang}, H. -Y. Karen and {Ruszkowski}, Mateusz and {Zweibel}, Ellen G.},
        title = "{Fermi and eROSITA bubbles as relics of the past activity of the Galaxy's central black hole}",
      journal = {Nature Astronomy},
         year = 2022,
        month = mar,
       volume = {6},
        pages = {584-591},
          doi = {10.1038/s41550-022-01618-x},
archivePrefix = {arXiv},
       eprint = {2203.02526},
 primaryClass = {astro-ph.HE},
       adsurl = {https://ui.adsabs.harvard.edu/abs/2022NatAs...6..584Y}
}

@ARTICLE{Gupta23,
       author = {{Gupta}, Anjali and {Mathur}, Smita and {Kingsbury}, Joshua and {Das}, Sanskriti and {Krongold}, Yair},
        title = "{Thermal and chemical properties of the eROSITA bubbles from Suzaku observations}",
      journal = {Nature Astronomy},
         year = 2023,
        month = jul,
       volume = {7},
        pages = {799-804},
          doi = {10.1038/s41550-023-01963-5},
archivePrefix = {arXiv},
       eprint = {2201.09915},
 primaryClass = {astro-ph.GA},
       adsurl = {https://ui.adsabs.harvard.edu/abs/2023NatAs...7..799G}
}

@ARTICLE{Porquet10,
       author = {{Porquet}, D. and {Dubau}, J. and {Grosso}, N.},
        title = "{He-like Ions as Practical Astrophysical Plasma Diagnostics: From Stellar Coronae to Active Galactic Nuclei}",
      journal = {Space Science Reviews},
         year = 2010,
        month = dec,
       volume = {157},
       number = {1-4},
        pages = {103-134},
          doi = {10.1007/s11214-010-9731-2},
archivePrefix = {arXiv},
       eprint = {1101.3184},
 primaryClass = {astro-ph.HE},
       adsurl = {https://ui.adsabs.harvard.edu/abs/2010SSRv..157..103P}
}

@ARTICLE{Mao19,
       author = {{Mao}, Junjie and {Kaastra}, Jelle S. and {Guainazzi}, Matteo and {Gonz{\'a}lez-Riestra}, Rosario and {Santos-Lle{\'o}}, Maria and {Kretschmar}, Peter and {Grinberg}, Victoria and {Kalfountzou}, Eleni and {Ibarra}, Aitor and {Matzeu}, Gabi and {Parker}, Michael and {Rodr{\'\i}guez-Pascual}, Pedro},
        title = "{CIELO-RGS: a catalog of soft X-ray ionized emission lines}",
      journal = {Astronomy and Astrophysics},
         year = 2019,
        month = may,
       volume = {625},
          eid = {A122},
        pages = {A122},
          doi = {10.1051/0004-6361/201935368},
archivePrefix = {arXiv},
       eprint = {1904.05446},
 primaryClass = {astro-ph.HE},
       adsurl = {https://ui.adsabs.harvard.edu/abs/2019A&A...625A.122M}
}

@ARTICLE{Mao22,
       author = {{Mao}, Junjie and {Del Zanna}, G. and {Gu}, Liyi and {Zhang}, C.~Y. and {Badnell}, N.~R.},
        title = "{R-matrix Electron-impact Excitation Data for the H- and He-like Ions with Z = 6-30}",
      journal = {Astrophysical Journal, Supplement},
         year = 2022,
        month = dec,
       volume = {263},
       number = {2},
          eid = {35},
        pages = {35},
          doi = {10.3847/1538-4365/ac9c57},
archivePrefix = {arXiv},
       eprint = {2210.13427},
 primaryClass = {physics.atom-ph},
       adsurl = {https://ui.adsabs.harvard.edu/abs/2022ApJS..263...35M}
}

@CONFERENCE{SIMIT_TDM,
  author  = {Yining Zheng and Yingyu Chen and Yuanxing Xu and Huanqian Zhang and Lihong Tang and Liliang Ying and Hangxing Xie and Bo Gao and Zhen Wang},
  title   = "{Development of sensor SQUID and Zappe interferometer switch for HUBS}",
  booktitle = {Proc. SPIE 12181, Space Telescopes and Instrumentation 2022: Ultraviolet to Gamma Ray}, 
  year    = {2022},
  volume = {12181},
  pages   = {121816I}
}

@ARTICLE{Cui20,
       author = {{Cui}, W. and {Chen}, L. -B. and {Gao}, B. and {Guo}, F. -L. and {Jin}, H. and {Wang}, G. -L. and {Wang}, L. and {Wang}, J. -J. and {Wang}, W. and {Wang}, Z. -S. and {Wang}, Z. and {Yuan}, F. and {Zhang}, W.},
        title = "{HUBS: Hot Universe Baryon Surveyor}",
      journal = {Journal of Low Temperature Physics},
         year = 2020,
        month = jan,
       volume = {199},
       number = {1-2},
        pages = {502-509},
          doi = {10.1007/s10909-019-02279-3},
       adsurl = {https://ui.adsabs.harvard.edu/abs/2020JLTP..199..502C}
}

@ARTICLE{Kuntz19,
       author = {{Kuntz}, K.~D.},
        title = "{Solar wind charge exchange: an astrophysical nuisance}",
      journal = {Astronomy and Astrophysicsr},
         year = 2019,
        month = jan,
       volume = {27},
       number = {1},
          eid = {1},
        pages = {1},
          doi = {10.1007/s00159-018-0114-0},
archivePrefix = {arXiv},
       eprint = {1811.06454},
 primaryClass = {astro-ph.HE},
       adsurl = {https://ui.adsabs.harvard.edu/abs/2019A&ARv..27....1K}
}

@ARTICLE{Gu23,
       author = {{Gu}, Liyi and {Shah}, Chintan},
        title = "{Charge exchange in X-ray astrophysics}",
      journal = {arXiv e-prints},
         year = 2023,
        month = jan,
          eid = {arXiv:2301.11335},
        pages = {arXiv:2301.11335},
          doi = {10.48550/arXiv.2301.11335},
archivePrefix = {arXiv},
       eprint = {2301.11335},
 primaryClass = {astro-ph.IM},
       adsurl = {https://ui.adsabs.harvard.edu/abs/2023arXiv230111335G}
}

@ARTICLE{Sanders2001,
       author = {{Sanders}, W.~T. and {Edgar}, Richard J. and {Kraushaar}, W.~L. and {McCammon}, D. and {Morgenthaler}, J.~P.},
        title = "{Spectra of the 1/4 keV X-Ray Diffuse Background from the Diffuse X-Ray Spectrometer Experiment}",
      journal = {Astrophysical Journal},
         year = 2001,
        month = jun,
       volume = {554},
       number = {2},
        pages = {694-709},
          doi = {10.1086/321424},
       adsurl = {https://ui.adsabs.harvard.edu/abs/2001ApJ...554..694S}
}

@ARTICLE{Rogantini22,
       author = {{Rogantini}, D. and {Costantini}, E. and {Zeegers}, S.~T. and {Mehdipour}, M. and {Psaradaki}, I. and {Raassen}, A.~J.~J. and {de Vries}, C.~P. and {Waters}, L.~B.~F.~M.},
        title = "{Magnesium and silicon in interstellar dust: X-ray overview}",
      journal = {Astronomy and Astrophysics},
         year = 2020,
        month = sep,
       volume = {641},
          eid = {A149},
        pages = {A149},
          doi = {10.1051/0004-6361/201936805},
archivePrefix = {arXiv},
       eprint = {2007.03329},
 primaryClass = {astro-ph.HE},
       adsurl = {https://ui.adsabs.harvard.edu/abs/2020A&A...641A.149R}
}

@ARTICLE{Psaradaki23,
       author = {{Psaradaki}, I. and {Costantini}, E. and {Rogantini}, D. and {Mehdipour}, M. and {Corrales}, L. and {Zeegers}, S.~T. and {de Groot}, F. and {den Herder}, J.~W.~A. and {Mutschke}, H. and {Trasobares}, S. and {de Vries}, C.~P. and {Waters}, L.~B.~F.~M.},
        title = "{Oxygen and iron in interstellar dust: An X-ray investigation}",
      journal = {Astronomy and Astrophysics},
         year = 2023,
        month = feb,
       volume = {670},
          eid = {A30},
        pages = {A30},
          doi = {10.1051/0004-6361/202244110},
archivePrefix = {arXiv},
       eprint = {2210.05778},
 primaryClass = {astro-ph.HE},
       adsurl = {https://ui.adsabs.harvard.edu/abs/2023A&A...670A..30P}
}

@ARTICLE{Henley12,
       author = {{Henley}, David B. and {Shelton}, Robin L.},
        title = "{An XMM-Newton Survey of the Soft X-Ray Background. II. An All-Sky Catalog of Diffuse O VII and O VIII Emission Intensities}",
      journal = {Astrophysical Journal, Supplement},
         year = 2012,
        month = oct,
       volume = {202},
       number = {2},
          eid = {14},
        pages = {14},
          doi = {10.1088/0067-0049/202/2/14},
archivePrefix = {arXiv},
       eprint = {1208.4360},
 primaryClass = {astro-ph.GA},
       adsurl = {https://ui.adsabs.harvard.edu/abs/2012ApJS..202...14H}
}

@ARTICLE{Gupta12,
       author = {{Gupta}, A. and {Mathur}, S. and {Krongold}, Y. and {Nicastro}, F. and {Galeazzi}, M.},
        title = "{A Huge Reservoir of Ionized Gas around the Milky Way: Accounting for the Missing Mass?}",
      journal = {Astrophysical Journal, Letters},
         year = 2012,
        month = sep,
       volume = {756},
       number = {1},
          eid = {L8},
        pages = {L8},
          doi = {10.1088/2041-8205/756/1/L8},
archivePrefix = {arXiv},
       eprint = {1205.5037},
 primaryClass = {astro-ph.HE},
       adsurl = {https://ui.adsabs.harvard.edu/abs/2012ApJ...756L...8G}
}

@ARTICLE{McCammon02,
       author = {{McCammon}, D. and {Almy}, R. and {Apodaca}, E. and {Bergmann Tiest}, W. and {Cui}, W. and {Deiker}, S. and {Galeazzi}, M. and {Juda}, M. and {Lesser}, A. and {Mihara}, T. and {Morgenthaler}, J.~P. and {Sanders}, W.~T. and {Zhang}, J. and {Figueroa-Feliciano}, E. and {Kelley}, R.~L. and {Moseley}, S.~H. and {Mushotzky}, R.~F. and {Porter}, F.~S. and {Stahle}, C.~K. and {Szymkowiak}, A.~E.},
        title = "{A High Spectral Resolution Observation of the Soft X-Ray Diffuse Background with Thermal Detectors}",
      journal = {Astrophysical Journal},
         year = 2002,
        month = sep,
       volume = {576},
       number = {1},
        pages = {188-203},
          doi = {10.1086/341727},
archivePrefix = {arXiv},
       eprint = {astro-ph/0205012},
 primaryClass = {astro-ph},
       adsurl = {https://ui.adsabs.harvard.edu/abs/2002ApJ...576..188M}
}

@ARTICLE{Das19,
       author = {{Das}, Sanskriti and {Mathur}, Smita and {Gupta}, Anjali and {Nicastro}, Fabrizio and {Krongold}, Yair},
        title = "{Multiple Temperature Components of the Hot Circumgalactic Medium of the Milky Way}",
      journal = {Astrophysical Journal},
         year = 2019,
        month = dec,
       volume = {887},
       number = {2},
          eid = {257},
        pages = {257},
          doi = {10.3847/1538-4357/ab5846},
archivePrefix = {arXiv},
       eprint = {1909.06688},
 primaryClass = {astro-ph.GA},
       adsurl = {https://ui.adsabs.harvard.edu/abs/2019ApJ...887..257D}
}

@INPROCEEDINGS{Wulf19,
       author = {{Wulf}, Dallas and {Eckart}, Mega E. and {Galeazzi}, Massimiliano and {Jaeckel}, Felix and {Kelley}, Richard L. and {Kilbourne}, Caroline A. and {McCammon}, Dan and {Morgan}, Kelsey M. and {Porter}, Frederick S. and {Szymkowiak}, Andrew E.},
        title = "{High Spectral Resolution Observation of the Soft Diffuse X-ray Background in the Direction of the Galactic Anti-Center}",
    booktitle = {American Astronomical Society Meeting Abstracts \#231},
         year = 2018,
       series = {American Astronomical Society Meeting Abstracts},
       volume = {231},
        month = jan,
          eid = {331.06},
        pages = {331.06},
       adsurl = {https://ui.adsabs.harvard.edu/abs/2018AAS...23133106W}
}

@ARTICLE{Katsuda11,
       author = {{Katsuda}, Satoru and {Tsunemi}, Hiroshi and {Mori}, Koji and {Uchida}, Hiroyuki and {Kosugi}, Hiroko and {Kimura}, Masashi and {Nakajima}, Hiroshi and {Takakura}, Satoru and {Petre}, Robert and {Hewitt}, John W. and {Yamaguchi}, Hiroya},
        title = "{Possible Charge-exchange X-ray Emission in the Cygnus Loop Detected with Suzaku}",
      journal = {Astrophysical Journal},
         year = 2011,
        month = mar,
       volume = {730},
       number = {1},
          eid = {24},
        pages = {24},
          doi = {10.1088/0004-637X/730/1/24},
archivePrefix = {arXiv},
       eprint = {1103.1669},
 primaryClass = {astro-ph.HE},
       adsurl = {https://ui.adsabs.harvard.edu/abs/2011ApJ...730...24K}
}

@ARTICLE{Uchida19,
       author = {{Uchida}, H. and {Katsuda}, S. and {Tsunemi}, H. and {Mori}, K. and {Gu}, L. and {Cumbee}, R.~S. and {Petre}, R. and {Tanaka}, T.},
        title = "{High Forbidden-to-resonance Line Ratio of O VII Discovered from the Cygnus Loop}",
      journal = {Astrophysical Journal},
         year = 2019,
        month = feb,
       volume = {871},
       number = {2},
          eid = {234},
        pages = {234},
          doi = {10.3847/1538-4357/aaf8a6},
archivePrefix = {arXiv},
       eprint = {1812.06616},
 primaryClass = {astro-ph.HE},
       adsurl = {https://ui.adsabs.harvard.edu/abs/2019ApJ...871..234U}
}

@ARTICLE{Zucker22,
       author = {{Zucker}, Catherine and {Goodman}, Alyssa A. and {Alves}, Jo{\~a}o and {Bialy}, Shmuel and {Foley}, Michael and {Speagle}, Joshua S. and {Gro{\^I}{\texttwosuperior}schedl}, Josefa and {Finkbeiner}, Douglas P. and {Burkert}, Andreas and {Khimey}, Diana and {Swiggum}, Cameren},
        title = "{Star formation near the Sun is driven by expansion of the Local Bubble}",
      journal = {Nature},
         year = 2022,
        month = jan,
       volume = {601},
       number = {7893},
        pages = {334-337},
          doi = {10.1038/s41586-021-04286-5},
archivePrefix = {arXiv},
       eprint = {2201.05124},
 primaryClass = {astro-ph.GA},
       adsurl = {https://ui.adsabs.harvard.edu/abs/2022Natur.601..334Z}
}

@article{WANG2022100027,
title = {Development of superconducting microcalorimeters for the {HUBS} mission},
journal = {Superconductivity},
volume = {4},
pages = {100027},
year = {2022},
issn = {2772-8307},
doi = {https://doi.org/10.1016/j.supcon.2022.100027},
url = {https://www.sciencedirect.com/science/article/pii/S2772830722000266},
author = {Sifan Wang and Guole Wang and Naihui Chen and Yanling Chen and Wei Cui and Jiao Ding and Fajun Li and Yajie Liang and Qian Wang and Yeru Wang}
}

@article{JIANG2023,
title = {Development of adiabatic demagnetization refrigerator for the HUBS mission},
journal = {Science Bulletin},
year = {2023},
issn = {2095-9273},
doi = {https://doi.org/10.1016/j.scib.2023.09.031},
url = {https://www.sciencedirect.com/science/article/pii/S2095927323006667},
author = {Chunyang Jiang and Chengzhe Li and Hai Jin and Wei Cui}
}

@ARTICLE{2023YangB,
       author = {{Yang}, Biao and {Gao}, ZhaoZhao and {Chen}, LiuBiao and {Liu}, SiXue and {Zhou}, Qiang and {Guo}, Jia and {Cui}, Chen and {Zhu}, WenXiu and {Jin}, Hai and {Zhou}, Yuan and {Wang}, JunJie},
        title = "{First high-frequency pulse tube cryocooler down to 2.5 K and its promising application in China's deep space exploration}",
      journal = {Science in China E: Technological Sciences},
         year = 2023,
        month = aug,
       volume = {66},
       number = {8},
        pages = {2454-2456},
          doi = {10.1007/s11431-022-2398-1},
       adsurl = {https://ui.adsabs.harvard.edu/abs/2023ScChE..66.2454Y}
}

@ARTICLE{LJJ_DetAss,
       author = {{Liu}, Jiejia. and {Wang}, Sifan and {Jin}, Hai. and {Cui}, Wei },
        title = "{Preliminary Design of Detector Assembly for
DIXE}",
      journal = {Journal of Low Temperature Physics},
         year = 2023,
        
       volume = {in special issue},
       number = { },
        pages = { },
         
       adsurl = {https://ui.adsabs.harvard.edu/abs/2020JLTP..199..502C}
}

@ARTICLE{SIMIT_LTD,
       author = {Yuanxing Xu and Yingyu Chen and Chaoqun Wang and Taiyu Li and Lihong Tang and Huanqian Zhang and Hangxing Xie and Bo Gao and Zhen Wang},
        title = "{Development of time division multiplexing technique for HUBS}",
      journal = {Journal of Low Temperature Physics},
         year = 2023,
        
       volume = {in special issue},
       number = { },
        pages = { },
}
\end{refcontext}
\end{document}